\documentclass[aps,twocolumn,showpacs]{revtex4}

\usepackage{graphicx}
\usepackage{amsmath}
\usepackage{color}
\begin{document}

\title{Theory of high energy optical conductivity and the role of oxygens in manganites}

\author{ Muhammad Aziz Majidi$^{1,4}$, Haibin Su$^{5}$,  Yuan Ping Feng$^{1}$,
 Michael R\"{u}bhausen$^{3,1}$, and  Andrivo Rusydi$^{1,2,3}$}
\email{phyandri@nus.edu.sg}
\affiliation{$^{1}$
NUSNNI-NanoCore, Department of Physics, Faculty of Science,
National University of Singapore, Singapore 117542, Singapore,}
\affiliation{$^{2}$
Singapore Synchrotron Light Source, National University of Singapore, Singapore 117603,
Singapore}
\affiliation{$^{3}$
Institut f\"{u}r Angewandte Physik, Universit\"{a}t Hamburg, Jungiusstrae 11,
D-20355 Hamburg, Germany. \\
Center for Free Electron Laser Science (CFEL), Notkestra{\ss}e 85,
D- 22607 Hamburg, Germany,}
\affiliation{$^{4}$Departemen Fisika, FMIPA, Universitas Indonesia,
Depok 16424, Indonesia}
\affiliation{$^{5}$
Division of Materials Science, Nanyang Technological University,
50 Nanyang Avenue, Singapore 639798, Singapore.}
\date{\today}

\begin{abstract}
  Recent experimental study reveals the optical conductivity
  of La$_{1-x}$Ca$_x$MnO$_3$ over a wide range of energy
  and the occurrence of spectral weight transfer as the system transforms
  from paramagnetic insulating to ferromagnetic metallic phase
  [Rusydi {\it et al.}, Phys. Rev. B {\bf 78}, 125110 (2008)].
  We propose a model and calculation within the Dynamical Mean Field Theory
  to explain this phenomenon. We find the role of oxygens in mediating
  the hopping of electrons between manganeses as the key that determines the
  structures of the optical conductivity.
  In addition, by parametrizing the hopping integrals through magnetization, our result suggests a possible scenario that explains the occurrence of spectral weight transfer, in which the ferromagnatic ordering increases the rate of electron transfer from O$_{2p}$ orbitals to upper Mn$_{e_g}$ orbitals while simultaneously decreasing the rate of electron transfer from O$_{2p}$ orbitals to lower Mn$_{e_g}$orbitals, as temperature is varied across the ferromagnetic transition.
  With this scenario, our optical conductivity calculation shows very good quantitative agreement with the experimental data.
\end{abstract}


\maketitle

{\em{Introduction.}}
Manganites have been the subject of extensive studies
since they have exhibited a wealth of fascitaning phenomena
such as the colossal magnetoresistance
(CMR), charge-, spin-, and orbital orderings, and transition from paramagnetic
insulator to ferromagnetic metal, as well as multiferroic behavior
\cite{Jin-Sc1994,Salamon-RMP2001,Saitoh-Nat2001,Cheong-NM2007}.
Upon hole doping, the transition from antiferromagnetic insulator
to ferromagnetic metal has been
argued to occur through a mixed-phase process \cite{Moreo-Science1999}.
Whereas for a fixed hole doping where ferromagnetic order is found,
insulator to metal transition simultaneously occurs as temperature
is lowered across the ferromagnetic transition \cite{Nucara-PRB2003}.
It has been generally assumed and experimentally confirmed
that the magnetic order in these
systems is driven by the double-exchange interactions
\cite{Anderson-PR1955,deGennes-PR1960,Quijada-PRB1998,Moreo-Science1999,
Amelitchev-PRB2001}.
However, explanation on other phenomena accompanying the ferromagnetic
transition seems to be far from complete, and remains as an open subject.

Several theories on the insulator-metal (I-M) transition accompanying
the ferromagnetic transition have been proposed
\cite{Millis-PRB1996,Ramakrishnan-PRL2004}.
Although the details of the models and scenarios of the I-M
transition proposed by these theories are quite different,
they have similar idea suggesting that the Jahn-Teller distortion along
with the electron-phonon interactions stabilize the insulating phase at high
temperatures, which is broken by the ferromagnetic order
below its transition temperatures.
These theories, however, have only addressed the static properties or
low energy phenomena, as their models implicitly assume that low energy
phenomena occuring in these materials are insensitive to possible
high energy excitations.
Many such models
\cite{Millis-PRL1996,Cepas-PRL2005,Lee-PRB2007,Stier-PRB2007,Rong-PRB2008,
Lin-PRB2008}
typically consider only effective hoppings between
Mn sites, while ignoring the electronic states in oxygen sites.
On the other hand, models that included local interactions and hybridization
in correlated materials, might expect pronounced
effects at higher energies that are connected to
charge-transfer or Mott-Hubbard physics \cite{Zaanen-PRL1985, Meinders-PRB1993, Yin-PRL2006, Phillips-RMP2010}.
Thus, the validity of such theories
may have to be tested through experimental studies on the band structures and
the optical properties over a wide range of energy.
In that respect, experimental studies of optical conductivity of manganites
as functions of temperature and doping in a much wider energy range
become crucial.

A recent study of optical conductivity by Rusydi {\it et al.}
\cite{Rusydi-PRB2008},
has revealed for the first time strong temperature and doping dependences in
La$_{1-x}$Ca$_x$MnO$_3$ for $x$ = 0.3 and 0.2.
The occurrence of spectral weight transfer has been strikingly found between
low ($<$3eV), medium (3-12eV),
and high energies ($>$12eV) across I-M transition.
In fact, as the temperature is decreased, the spectral weight transfer appears
more noticeably in the medium and high energy regions than it does in
the low energy region.
Observing how the spectral weight in each region of energy simultaneously
changes as temperature is decreased passing the ferromagnetic transition
temperature ($T_{FM}$),
one may suspect that there is an interplay between low, medium, and
high energy charge transfers that may drive many phenomena occuring
in manganites, including the I-M transition.
This conjecture is related to the fact that the hopping of an electron
from one Mn site to another Mn site can only occur through an O site.

Considering the difference between the on-site energy
of the manganese and that of the oxygen that could be about 5-8 eV
\cite{Picket-PRB1996},
the Mn-O hoppings occur with high energy transfer.
We hypothesize that if such high energy hoppings can mediate a ferromagnetic
order, then other low or high energy phenomena may possibly occur
simulataneously. Thus, the mechanism of I-M transition
in the dc conductivity may not be completely separated from what
appears as the decrease (increase) of the spectral weight in the
medium (high) energy region of the optical conductivity,
all of which together may be driven by the ferromagnetic ordering.
Theories based on effective low energy models which only consider
Mn sites while ignoring O sites would not be able to address this.

Motivated by the aforementioned conjecture, we develop a simple
but more general model, in which oxygens are explicitly incorporated.
In this paper, we propose our model and calculation of
the optical conductivity of La$_{1-x}$Ca$_x$MnO$_3$ within the
Dynamical Mean Field Theory, to explain the experimental results
of Ref. \cite{Rusydi-PRB2008}. Our calculated optical conductivity shows
that both oxygens and mangeneses play important roles in forming
structures similar to those of the experimental results.
Further, with some additional argument,
our calculation captures qualitatively correctly the temperature dependence
of the optical conductivity as the system transforms from paramagnetic
to ferromagnetic phase.

\begin{figure}
\includegraphics[width=2.0in]{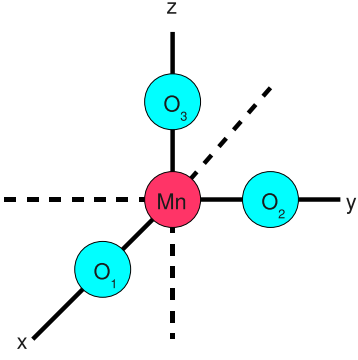}
\caption{(Color online) Simplified crystal structure of the model.
 The crystal structure is assummed cubic with the unit cell
containing only one Mn and three O atoms. The O atoms are labeled 1,2, and 3
 to distinguish the $p-$orbitals belonging to different O atoms used
to construct our Hamiltonian.}
\label{model_structure}
\end{figure}

{\em{Model.}}
As shown in Fig. \ref{model_structure}, we model the crystal structure
of La$_{1-x}$Ca$_x$MnO$_3$ such that each unit cell forms a cube with lattice constant $a$ set equal to 1, and
contains only one Mn and three O sites, thus ignoring the presence of
La and Ca atoms that we believe not to contribute much to
the structures and temperature dependence of the optical conductivity.
We choose 10 basis orbitals to construct our Hilbert space, which we order as
the following: $|{\rm Mn} ~ e_{g~ upper,\uparrow}\rangle$,
$|{\rm Mn} ~ e_{g~ lower,\uparrow}\rangle$, $|{\rm O}_1 ~ p_\uparrow\rangle$,
$|{\rm O}_2 ~ p_\uparrow\rangle$, $|{\rm O}_3 ~ p_\uparrow\rangle$,
 $|{\rm Mn} ~ e_{g~ upper,\downarrow}\rangle$,
$|{\rm Mn} ~ e_{g~ lower,\downarrow}\rangle$, $|{\rm O}_1 ~ p_\downarrow\rangle$,
$|{\rm O}_2 ~ p_\downarrow\rangle$, and
$|{\rm O}_3 ~ p_\downarrow\rangle$. Note that the distinction between
$|~ e_{g~ upper}\rangle$ and $|~ e_{g~ lower}\rangle$ states is associated with
the Jahn-Teller splitting. Using this set of bases we propose a Hamiltonian:
\begin{eqnarray}
\label{eq:Hamiltonian}
 H&=&\frac{1}{N}\sum_{\bf k} \eta^\dagger_{\bf k} [H_0({\bf k})]\eta_{\bf k}
+ \sum_{i,\sigma,\sigma'}U n_{u_i \sigma}n_{l_i \sigma'}
\nonumber \\
&& + \sum_{i \sigma}U_u n_{u_i \uparrow}n_{u_i \downarrow}
+ \sum_{i \sigma}U_l n_{l_i \uparrow}n_{l_i \downarrow} \nonumber \\
&& - \sum_i J_H {\bf S}_i{\bf .s}_i .
\end{eqnarray}
\noindent The first term in the Hamiltonian
is the kinetic part, whereof $\eta^\dagger_{\bf k}$
is a row vector whose elements are the creation operators associated with the 10
basis orbitals, and $\eta_{\bf k}$  is its hermitian conjugate
containing the corresponding destruction operators.
Here we consider that each Mn site contributes 4 $e_g$ orbitals
(upper and lower each of which is with up and down spins),
and the three O sites contribute 6 orbitals (3 from each site each of which
is with up and down spins).
$[H_0({\bf k})]$ is a 10$\times$10 matrix in momentum space whose structure
is arranged in four 5$\times$5 blocks
corresponding to their spin directions as
\begin{eqnarray}
\label{eq:H0}
 [H_0({\bf k})]=
\left[
\begin{matrix}
H_0({\bf k})_\uparrow&{\bf O}\\
{\bf O} &H_0({\bf k})_\downarrow
\end{matrix}
\right],
\end{eqnarray}
where ${\bf O}$ is a zero matrix of size 5$\times$5, and (referring to
the choice of coordinates in Fig. \ref{model_structure})
\begin{widetext}
\begin{equation}
\label{eq:H0_5x5}
H_0({\bf k})_{\uparrow(\downarrow)} =
\left[
\begin{smallmatrix}
E_{JT} & 0 & t_{Mn-O}^{(1)}(1+e^{-ik_x}) &  t_{Mn-O}^{(1)}(1+e^{-ik_y})
& t_{Mn-O}^{(1)}(1+e^{-ik_z}) \\
0 & -E_{JT} &  t_{Mn-O}^{(2)}(1+e^{-ik_x}) &  t_{Mn-O}^{(2)}(1+e^{-ik_y})
& t_{Mn-O}^{(2)}(1+e^{-ik_z}) \\
 t_{Mn-O}^{(1)}(1+e^{ik_x}) &  t_{Mn-O}^{(2)}(1+e^{ik_x}) & E_p &
t_{O-O}(1+2e^{ik_x}+2e^{-ik_y})
& t_{O-O}(1+2e^{ik_x}+2e^{-ik_z})\\
 t_{Mn-O}^{(1)}(1+e^{ik_y}) &  t_{Mn-O}^{(2)}(1+e^{ik_y}) &
t_{O-O}(1+2e^{-ik_x}+2e^{ik_y}) & E_p
& t_{O-O}(1+2e^{ik_y}+2e^{-ik_z}) \\
 t_{Mn-O}^{(1)}(1+e^{ik_z}) &  t_{Mn-O}^{(2)}(1+e^{ik_z}) &
t_{O-O}(1+2e^{-ik_x}+2e^{ik_z}) &t_{O-O}(1+2e^{-ik_x}+2e^{ik_z})
& E_p
\end{smallmatrix}
\right].
\end{equation}

\end{widetext}

\noindent The diagonal elements of $H_0({\bf k})_{\uparrow(\downarrow)}$
represent the local energies, while the off-diagonal elements
represent the hybridizations between orbitals. The first two diagonal elements,
i.e. $E_{JT}$ and $-E_{JT}$, correspond to the Mn$_{e_g}$ orbital energies
which are split due to the presumedly static Jahn-Teller distortion.
Each of the remaining three diagonal elements, i.e. $E_p$, corresponds to the
local energy of the O$_{2p}$ orbital. The parameter $t_{Mn-O}^{(1)}$
($t_{Mn-O}^{(2)}$) corresponds to hopping between the upper (lower) Mn$_{e_g}$
orbital and the nearest O$_{2p}$ orbital.
Whereas $t_{O-O}$ corresponds to hopping between nearest O$_{2p}$ orbitals.

The second term in Eq. (\ref{eq:Hamiltonian}) represents the Coulomb repulsions
between the upper and lower Mn$_{e_g}$ orbitals in a site.
The third and forth terms represent the intra-orbital Coulomb repulsions.
In this work, we take $U_u$ and $U_l$ to be infinity, forbidding
double occupancy in each of the lower and upper Mn$_{e_g}$ orbitals.
Finally, the fifth term represents the double-exchange magnetic interactions
between the local spins of Mn, ${\bf S}$, formed by the strong Hund's coupling
among three $t_{2g}$ electrons giving {\it S}=3/2,
and the itinerant spins of the upper and lower
Mn$_{e_g}$ electrons, ${\bf s}$.
Note that we use a well-accepted general assumption that
the on-site Coulomb repulsion
in each $t_{2g}$ orbital and the Hund's coupling among the $t_{2g}$ orbitals
are so strong to keep the occupancy of the three $t_{2g}$ levels
fixed at high spin configuration.
Thus the charge degrees of freedom of the three  $t_{2g}$ electrons
become frozen, and the remaining
degree of freedom to be considered is the orientation of
the collective spin 3/2.

{\em{Method.}}
To solve our model, we use the Dynamical Mean Field Theory
\cite{Georges-RMP1996}.
First, we define the Green function of the system, which is a $10\times10$
matrix,
\begin{eqnarray}
\label{eq:GF_k}
[G({\bf k},z)] &=& \biggr[ [H_0({\bf k})] + [\Sigma(z)] \biggr]^{-1},
\end{eqnarray}
with $z$ the frequency variable.
Then, we coarse-grain it over the Brillouin zone as
\begin{eqnarray}
\label{eq:GF}
[G(z)]&=&\frac{1}{N}\sum_{\bf k} [G({\bf k},z)].
\end{eqnarray}

In defining $[G({\bf k},z)]$, all the interaction parts of the Hamiltonian
(all terms other than the kinetic part), are absorbed into a
momentum-independent self energy matrix,
$[\Sigma(z)]$, which will be solved self-consistently.
Note that in this algorithm, we need to go over the self-consistent loops
in both Matsubara ($z=i\omega_n+\mu$) and real frequency ($z=\omega+i0^+$).

On taking $U_u$ and $U_l$ to be infinity, to some approximation,
we forbid the double occupancies in states
$|{\rm Mn} ~ e_{g~ upper,\uparrow}\rangle$,
$|{\rm Mn} ~ e_{g~ lower,\uparrow}\rangle$,
 $|{\rm Mn} ~ e_{g~ upper,\downarrow}\rangle$, and
$|{\rm Mn} ~ e_{g~ lower,\downarrow}\rangle$
by throwing them out of our Hilbert space.
To do this, according to the structure of Hamiltonian matrix
in Eqs. (\ref{eq:H0}) and (\ref{eq:H0_5x5}),
we multiply the weights of all the diagonal elements with indices 1, 2, 6,
and 7, and all the corresponding off-diagonal
elements connecting any pair of them by a half.
Thus, after obtaining the matrix $[G(z)]$ from Eq. (\ref{eq:GF}),
the effective $[G(z)]$ (let's call it  $[G(z)]_{\rm eff}$) can be obtained
by multiplying each of the following blocks of  $[G(z)]$ by a half,
while keeping the remaining elements unchanged, that is
\begin{eqnarray}
\label{eq:restriction}
\left[
\begin{matrix}
G_{11}&G_{12}\\
G_{21}&G_{22}\\
\end{matrix}
\right] &\Rightarrow&
\frac{1}{2}
\left[
\begin{matrix}
G_{11}&G_{12}\\
G_{21}&G_{22}\\
\end{matrix}
\right], \nonumber \\
\left[
\begin{matrix}
G_{16}&G_{17}\\
G_{26}&G_{27}\\
\end{matrix}
\right] &\Rightarrow&
\frac{1}{2}
\left[
\begin{matrix}
G_{16}&G_{17}\\
G_{26}&G_{27}\\
\end{matrix}
\right], \nonumber \\
\left[
\begin{matrix}
G_{61}&G_{62}\\
G_{71}&G_{72}\\
\end{matrix}
\right] &\Rightarrow&
\frac{1}{2}
\left[
\begin{matrix}
G_{61}&G_{62}\\
G_{71}&G_{72}\\
\end{matrix}
\right], \nonumber \\
\left[
\begin{matrix}
G_{66}&G_{67}\\
G_{76}&G_{77}\\
\end{matrix}
\right] &\Rightarrow&
\frac{1}{2}
\left[
\begin{matrix}
G_{66}&G_{67}\\
G_{76}&G_{77}\\
\end{matrix}
\right].
\end{eqnarray}
The ``mean-field'' Green function can then be extracted as
\begin{eqnarray}
\label{eq:GF_mf}
 [{\cal G}(z)]=\biggr[ [G(z)]_{\rm eff}^{-1}+[\Sigma(z)]
 \biggr]^{-1}.
\end{eqnarray}

Next, we construct the local self energy matrix, $[\Sigma_{n_l}(z)]$,
corresponding to the second and the fifth terms of the Hamiltonian. Here
$n_l\in\{0,1\}$ is the occupation number of the lower Mn$_{e_g}$ orbital.
The elements of the $10\times10$ matrix $[\Sigma_{n_l}(z)]$ are all zero except
for the blocks
\begin{eqnarray}
\left[
\begin{matrix}
\Sigma_{11}&\Sigma_{16}\\
\Sigma_{61}&\Sigma_{66}\\
\end{matrix}
\right] &=&
\left[
\begin{smallmatrix}
-J_HS\cos\theta+n_lU& -J_HS(\cos\phi+i\sin\phi)\\
-J_HS(\cos\phi-i\sin\phi)&J_HS\cos\theta+n_lU
\end{smallmatrix}
\right], \nonumber \\
\left[
\begin{matrix}
\Sigma_{22}&\Sigma_{27}\\
\Sigma_{72}&\Sigma_{77}\\
\end{matrix}
\right] &=&
\left[
\begin{smallmatrix}
-J_HS\cos\theta & -J_HS(\cos\phi+i\sin\phi)\\
-J_HS(\cos\phi-i\sin\phi)&J_HS\cos\theta
\end{smallmatrix}
\right].\nonumber \\
\end{eqnarray}
The local interacting Green function matrix is then calculated through
\begin{eqnarray}
\label{eq:GF_loc}
[G_{n_l}(z)]=\biggr[ [{\cal G}(z)]^{-1}-[\Sigma_{n_l}(z)]
\biggr]^{-1},
\end{eqnarray}
where $\theta$ and $\phi$ are the corresponding angles representing the
direction of ${\bf S}$ in the spherical coordinate.

For each Mn site with a given $n_l$, the probability of Mn spin
${\bf S}$ having a direction with angle $\theta$ with respect to
the direction of magnetization (which is defined as the $z-$axis) is given by
\begin{eqnarray}
\label{eq:P_theta}
 P_{n_l}(\cos\theta)=\frac{e^{-S_{n_l}(\cos\theta)}}{Z_{n_l}},
\end{eqnarray}
where
\begin{eqnarray}
\label{eq:Z_nl}
Z_{n_l}=\int d(\cos\theta)e^{-S_{n_l}(\cos\theta)}
\end{eqnarray}
is the local partition function, and
\begin{eqnarray}
\label{eq:S_nl}
S_{n_l}(\cos\theta)=-\sum_n \ln \det [G_{n_l}(i\omega_n)]e^{-i\omega_n0^+}
\end{eqnarray}
is the effective action.

We need to average $[G_{n_l}(z)]$ over all possible
$\theta$ and $n_l$ values as
\begin{eqnarray}
\label{eq:GF_ave}
[G(z)]_{\rm ave}&=&( 1-\langle n_l \rangle ) \int d(\cos\theta) P_0(\cos\theta)
[G_0(z)] \nonumber \\
&&+ ~ \langle n_l \rangle \int d(\cos\theta) P_1(\cos\theta) [G_{1}(z)],
\end{eqnarray}
where $\langle n_l \rangle$ is the average occupation of lower Mn$_{e_g}$
orbital. The new self energy matrix is extracted through
\begin{eqnarray}
\label{eq:new_Sigma}
[\Sigma(z)]=[{\cal G}(z)]^{-1}-[G(z)]_{\rm ave}^{-1} ~ .
\end{eqnarray}
Finally, we feed this new self energy matrix back
into the definition of Green function in Eq. (\ref{eq:GF_k}),
and the iteration process continues until $[\Sigma(z)]$ converges.

After the self-consistency is achieved, we can compute the
density of states as
\begin{eqnarray}
\label{eq:DOS}
{\rm DOS}(\omega)=- \frac{1}{\pi}{\rm Im Tr}[G(\omega+i0^+)] ~ .
\end{eqnarray}
We can also compute the optical conductivity tensor as
\begin{eqnarray}
\label{eq:opt_cond}
\sigma_{\alpha \beta}(\omega) =
\frac{\pi e^2}{\hbar a d}
\int d\nu \biggr( \frac{f(\nu,T)-f(\nu+\omega,T)}{\omega} \biggr) \times
~~~~~~~~ \nonumber \\
\frac{1}{N} \sum_{\bf k}
~ {\rm Tr} [v_{\alpha}({\bf k})][A({\bf k},\nu)]
[v_{\beta}({\bf k})][A({\bf k},\nu+\omega)],
\nonumber \\
\end{eqnarray}
where
$[v_{\lambda}({\bf k})]=
\partial[H_0({\bf k})]/ \partial k_{\lambda}$
is the Cartesian component of the velocity matrix,
 $[A({\bf k},\nu)] = \big([G({\bf k},\omega+i0^+)]-
[G({\bf k},\omega-i0^+)]\big)/(2\pi i)$
the spectral function matrix, and
$f(\nu,T)$ the Fermi distribution function.
Note that the dimensional pre-factor $\pi e^2/(\hbar a d)$ \cite{Georges-RMP1996}  with $d=3$ is
introduced to restore the proper physical unit, since the rest of the expression was derived
by setting $e=\hbar=a=1$.
In our model, the system is isotropic, and we are only interested in
the tranverse components
$\sigma_{\alpha \alpha}(\omega)\equiv \sigma(\omega)$, which are equal
for all $\alpha \in \{x,y,z\}$.

\begin{figure}
\includegraphics[width=3.2in]{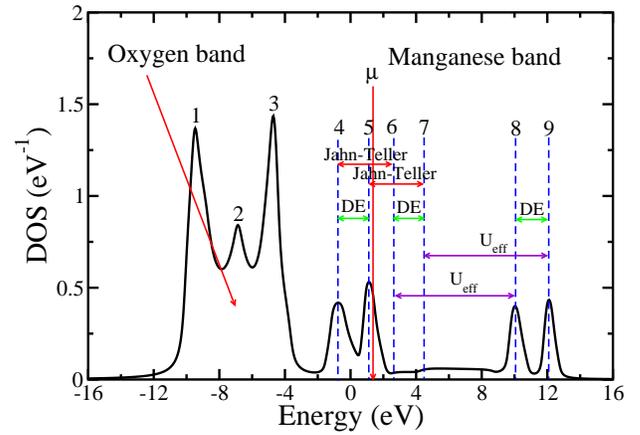}
\caption{(Color online) Calculated density of states (DOS).
See text for the parameter values used in this figure and the
detailed explanation of the structures of the DOS.
 }
\label{DOS}
\end{figure}

{\em{Results.}}
Our calculated density of states is shown in Fig.
\ref{DOS}. The parameter values used for this calculation
are $E_{JT}$=0.5eV, $E_p$=-6.5eV, $t_{Mn-O}^{(1)}$=1.2eV,
$t_{Mn-O}^{(2)}$=0.8eV, $t_{O-O}$=0.6eV, $U$=10eV, $J_H$=1.5eV, $a$=3.9345\AA, and
$T\approx194$K (corresponding to $\beta\equiv1/T=60$ eV$^{-1}$).
These parameter values are chosen considering rough estimates given in other
papers \cite{Millis-PRB1996,Ramakrishnan-PRL2004,Picket-PRB1996}
and adjusted so as to give best agreement with the
experimental optical conductivity data in Ref. \cite{Rusydi-PRB2008}.
The DOS is normalized such that the integrated area is equal to 8, since in
each unit cell there are 6 orbitals coming from oxygens
and effectively 2 from manganeses, considering the restriction given by
relation (\ref{eq:restriction}).
The chemical potential is self-consistently adjusted
to satisfy the electron filling of 6+(1-$x$)=6.7,
mimicking the situation of La$_{1-x}$Ca$_{x}$MnO$_3$ for $x=0.3$.
The structures of the density of states can be explained as the following.
The three peaks labeled 1,2,3 result from the fact that there are
three oxygen atoms in a unit cell, where the degeneracy is broken into three
levels by the hybridization between $2p$ orbitals of the neighboring oxygen
atoms. The structures labeled with 4 through 9
result from the $e_g$ orbitals of mangeneses.
As shown in the figure, there are three mechanisms that split the Mn$_{e_g}$
states into 6 levels:
static Jahn-Teller (JT) distortion, Coulomb repulsion U$_{\rm eff}$)
between lower and upper
JT-split $e_g$ states, the double-exchange (DE) interaction between spins of
electrons in the lower and upper JT-split states and the Mn spins formed
by the Hund's coupling among the Mn $t_{2g}$ electrons \cite{Note1}.

\begin{figure}
\includegraphics[width=3.2in]{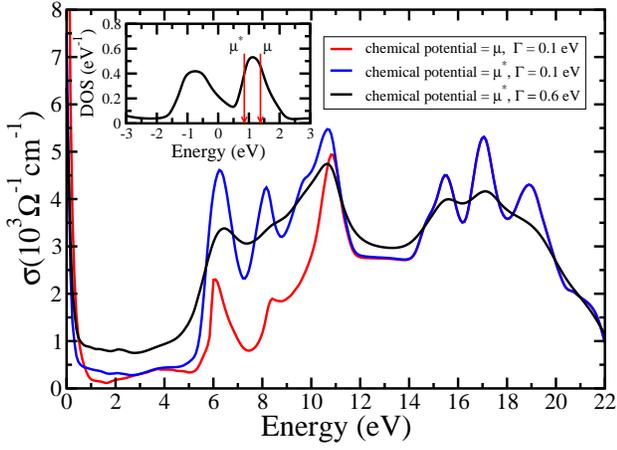}
\caption{(Color online) Calculated optical conductivity.
Main panel: The red and blue curves represent the calculated
optical conductivites for different position of chemical potential, $\mu$
and $\mu^*$, respectively, using only a small broadening ($\Gamma$=0.1 eV).
The black curve represents the result using chemical potential at $\mu^*$
with a bigger broadening ($\Gamma$=0.6eV).
Inset: Region in the density of states
showing how the position of chemical potential is shifted.
See text for detailed explanation.
}
\label{Cond}
\end{figure}

Figure  \ref{Cond} shows our calculated optical conductivity for $T\approx194$K
($>T_{FM}$). The ferromagnetic
transition temperature for this set of parameters is roughly
$T_{FM}\approx160$K (based on extrapolation of the mean-field trend).
The parameter values for $T\approx194$K are the same as those used in
Fig.  \ref{DOS}. In Figure \ref{Cond}, we demonstrate how we tune the
the profile of the optical conductivity to achieve the best resemblance
with the experimental data in Ref. \cite{Rusydi-PRB2008}.
It is important to note that our model is not meant to address the dc
conductivity, as we already anticipate that it cannot form an insulating
(or nearly insulating) phase at $T>T_{FM}$, possibly due to not incorporating
electron-phonon interactions
\cite{Millis-PRB1996,Ramakrishnan-PRL2004,Ederer-PRB2007}.
Rather, our goal is to show how this simple
model can capture qualitatively the general profile of the optical
conductivity from about 1 eV away from the Drude peak up to 22 eV
(the energy limit of the experimental data).

\begin{figure}
\includegraphics[width=3.2in]{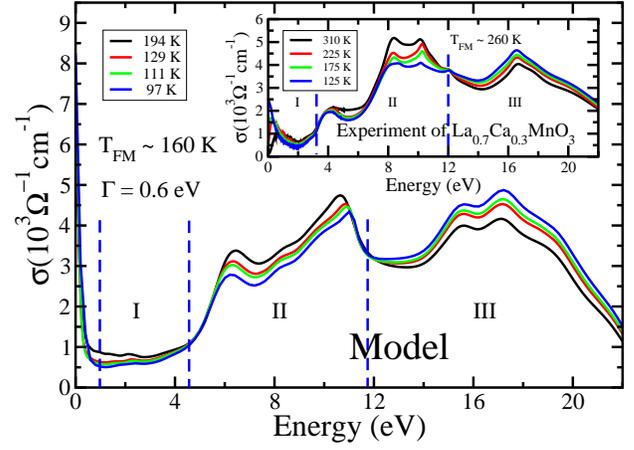}
\caption{(Color online) Spectral weight transfer in the optical conductivity.
Main panel: Results of the model.
Excluding the region containing the Drude peak (0-1eV),
the energy range is divided into three regions I,II, and III.
The black curve represents the optical
conductivity in the paramagnetic phase,
while the red, green, and blue curves
correspond successively to lower temperatures in the ferromagnetic phase
\cite{Note2}.
The borders between regions I-II
and II-III, denoted by the blue vertical dashed lines,
are defined such that the curves are crossing at these energies.
Inset: A replot of the corresponding experimental data from
Ref. \cite{Rusydi-PRB2008} for comparison.
}
\label{Cond_Tvaried}
\end{figure}

On calculating the optical conductivity from Eq. (\ref{eq:opt_cond})
we introduce an imaginary self energy for the O$_{2p}$
states, $-i\Gamma$, where $\tau=1/\Gamma$ corresponds to the lifetime
of the O$_{2p}$ states.
The red curve in Fig. \ref{Cond} shows the result if we use the
self-consistent chemical potential, $\mu$, in Eq. (\ref{eq:opt_cond}).
Here, we observe that the resulting profile around the medium energy
region ($\approx$ 5-11 eV) does not satisfactorily resembles that of the
experimental data in Ref. \cite{Rusydi-PRB2008}, since some spectral weight
seems to be missing in that region. We argue that the reason for this is related
to the fact that our self-consistent chemical potential, $\mu$, does not lie
inside a pseudogap as it probably would if we incorporate electron-phonon
interactions. In this model, we only have a pseudogap that
results from the double-exchange splitting,
where $\mu$ falls slightly to the right
outside of this pseudogap. To remedy the missing of spectral weight, we
slightly shift the position of chemical potential to the left, i.e.
from $\mu$ to $\mu^*$ as shown in
the inset of Fig. \ref{Cond}. Using this new chemical potential, $\mu^*$,
the resulting optical conductivity, shown in the blue curve,
resembles the experimental data better. This suggests that the
true chemical potential may actually lie inside a pseudogap similar to the
situation as though it lies at $\mu^*$.
(Note that, as long as considering the optical conductivity region
about 1 eV away from the Drude peak, choosing $\mu^*$ between 0.1 and 0.9 eV,
i.e. around the valley, lead to similar results.)
Although the profile of the blue curve is already better than the red one,
it still has more pronounced stuctures than the actual experimental data does.
To further tune the calculated
optical conductivity to better resemble the experimental data, we find that
the overly pronounced structures can be broadened by enlarging
the O$_{2p}$ imaginary self energy upto $\Gamma=0.6$ eV. The result after the
broadening, which is shown in the black curve, looks very similar
to the experimental results shown
in Fig. 2(b) of Ref. \cite{Rusydi-PRB2008} (replotted in the inset of
Fig. \ref{Cond_Tvaried}).
This similarity in both magnitude and profile of the energy
dependence may be a good measure of the validity of our model.

Now we discuss how the model captures the spectral weight transfer when
temperature is decreased from $T>T_{FM}$ to $T<T_{FM}$. First,
we divide the energy range into three regions:
I (low:$\approx$1-3eV), II (medium:$\approx$3-12eV), and III
(high:$\gtrsim$12eV), following the division made for the experimental
data in Ref. \cite{Rusydi-PRB2008}, except that we exclude the region around
the Drude peak from our discussion, since to obtain the correct values
of conductivity in that region requires a more accurate description of the
renormalized band structure around the chemical potential. If we decrease
temperature from the paramagnetic to ferromagnetic phase while keeping
all the parameters constant, we find no significicant change in the optical
conductivity, thus spectral weight transfer does not occur in this way.
If we inspect how Eq. (\ref{eq:opt_cond}) determines the optical conductivity,
we see that the change in optical conductivity may become more
significant if either the spectral function, $[A({\bf k},\nu)]$, or the
velocity operator, $[v_{\alpha}({\bf k})]$, changes significantly while
temperature changes. Within our model this can only be accomodated if we
allow some parameters to depend on temperature by some manner.
By comparing the structures of optical conductivity and the corresponding DOS profile,
it is clear that the spectral weight in the medium
energy region comes mostly from transitions from $O_{2p}$ to lower Mn$_{e_g}$
states, while in the high energy region from $O_{2p}$ to upper Mn$_{e_g}$
states. This fact may suggest that the hopping parameters
$t_{Mn-O}^{(1)}$ and  $t_{Mn-O}^{(2)}$ depend on temperature.
Furthermore, since the spectral weight transfer occurs most notably across
and below $T_{FM}$,
the temperature dependence of $t_{Mn-O}^{(1)}$ and  $t_{Mn-O}^{(2)}$ may be related to
spin correlation.

The actual interplay resulting in such a temperature dependence is believed to be
very complicated, since it may involve orbital effects on the dynamic electron-phonon coupling
and spin correlation. In that regard, our present model, which is not an {\it ab initio} based
model, cannot naturally capture these temperature effects.
Thus, to capture the plausible physics within our present model,
we turn to the phenomenological approach by
parametrizing the totally non-trivial temperature effecs on hopping integrals
through magnetization.
In the simplest level, we may assume a linear dependence of the hopping integrals
$t_{Mn-O}^{(1)}$ and
$t_{Mn-O}^{(2)}$ on the magnetization.
Hence, we may write

\begin{eqnarray}
\label{eq:t1_on_M}
t_{Mn-O}^{(1)}(M)=t_{Mn-O}^{(1)}(0)\biggr(1+c_1\frac{M}{M_s}\biggr), \\
\label{eq:t2_on_M}
t_{Mn-O}^{(2)}(M)=t_{Mn-O}^{(2)}(0)\biggr(1+c_2\frac{M}{M_s}\biggr),
\end{eqnarray}
where $M/M_s$ is the ratio of magnetization to the saturated magnetization,
and $c_1$, and $c_2$ are constants.

\begin{figure}
\includegraphics[width=3.2in]{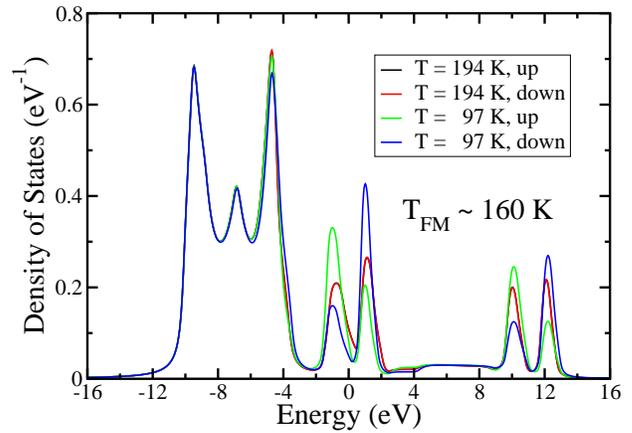}
\caption{(Color online) Spin-dependent density of states.
The black and the red curves lie on top of each other as the spin-up and
spin-down components of the DOS are identical in the paramagnetic phase.
While, the green and the blue curves look quite distinct as DOS becomes
polarized in the ferromagnetic phase.
 }
\label{DOS_plrzd}
\end{figure}

Using relations (\ref{eq:t1_on_M}) and (\ref{eq:t2_on_M}), taking
$t_{Mn-O}^{(1)}(0)$=1.2 eV, $t_{Mn-O}^{(2)}(0)$=0.8 eV,
$c_1\approx0.23$, and $c_2\approx$ -0.35,
at $T$=97K for which $M/M_s$=0.357, for instance, we obtain that
$t_{Mn-O}^{(1)}$ is enhanced to be $\approx$ 1.3 eV, while
$t_{Mn-O}^{(1)}$ is suppressed to be $\approx$ 0.7 eV.
The results for four different temperatures are shown in
Fig. \ref{Cond_Tvaried}. 
As shown in the main panel,
our calculation shows that the spectral weight simultaneously decreases
(increases) in the medium
(high) energy region of the optical conductivity as the system becomes
ferromagnetic \cite{Note3}.
Our calculation also produces a less noticeable
decrease of the spectral weight in the low energy region as observed
in the experimental data (see the inset).
In both the main panel and the inset,
the black curve represents the optical conductivity in the paramagnetic phase,
while the red, green, and blue curves correspond successively to lower
temperatures in the ferromagnetic phase.
If we define the positions of the
borders between energy regions I-II and II-III such that all the
curves are crossing at these energies, we obtain that theoretical
values of these energies are similar to the experimental ones.
Note that the temperatures varied in the theoretical and the
experimental results should not be compared quantitatively, since
the theoretical $T_{FM}$ is about 100 K too small compared to
the experimental one, possibly due to neglecting other possible exchange
interactions in our model. Despite this,
we believe that any improvement of $T_{FM}$ by such additional terms
would not change the physics presented in this paper.
To show the difference in the density of states between paramagnetic and
ferromagnetic phases,  we display
the spin dependent DOS for $T\approx194$K
and $T\approx97$K in Fig. \ref{DOS_plrzd}.

\begin{figure}
\includegraphics[width=3.2in]{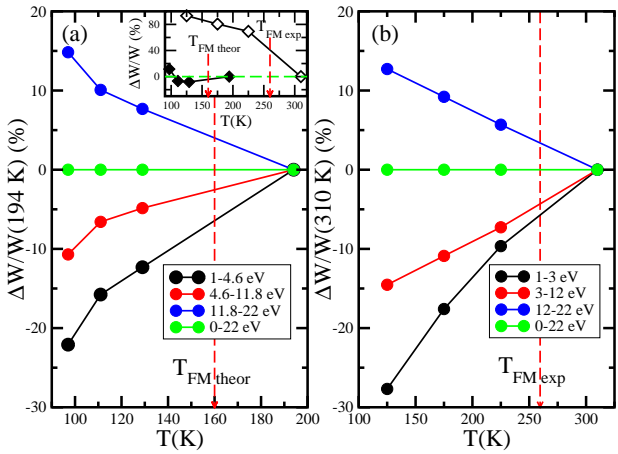}
\caption{
(Color online) Relative spectral-weight changes.
$\Delta W(T)/W$ from (a) our calculations,
and (b) the experimental results of
Ref. \cite{Rusydi-PRB2008}, for different regions of energy.
$\Delta W/W$ is defined as spectral-weight difference
$\Delta W = \int_{\omega_1}^{\omega_2} \big(\sigma(\omega,T)-
\sigma(\omega,T_{PM})\big)d\omega$ normalized to
$W(T_{PM}) = \int_{\omega_1}^{\omega_2} \sigma(\omega,T_{PM}) d\omega$,
where in this case $T_{PM}$ is 194 K in (a) and 310 K in (b).
Positions of the theoretical and the experimental $T_{FM}$ are
indicated by vertical red dashed lines in each panel.
Inset in (a) is
comparison between $\Delta W(T)/W$
 from the calculations (black filled diamonds)
and from the experiments of Ref. \cite{Rusydi-PRB2008} (black empty diamonds)
 for 0-1 eV region.
The horizontal green dashed line in the inset is just to highlight
the zero position of $\Delta W/W$.
}
\label{DeltaW_over_W}
\end{figure}

To demonstrate further how the spectral-weight transfers in our model compare
with the experimetal results, we display the relative spectral-weight changes
for different regions of energy in Fig. \ref{DeltaW_over_W}. Comparing
results in Fig. \ref{DeltaW_over_W}(a) and (b), it is clear that
for every region of energy, I (low), II (medium), and III (high),
(excluding 0-1 eV), our calculations give
exactly same trends as those shown by the experimental results. These
suggest that the ingredients incorporated in our model are adequate
to explain the occurrence of spectral-weight transfers in
La$_{1-x}$Ca$_x$MnO$_3$  in the energy range up to 22 eV.
In that respect, one may argue, for instance, that
the high-spin state ($S=3/2$) of the $t_{2g}$-electrons may become unstable
as the system is optically excited by high-energy photons.
Accordingly, transitions from high to low-spin states, or excitations
of electrons from  $t_{2g}$ to $e_g$ levels may occur.
Our present model does not incorporate those possibilities. However, our
calculations prove that the model is capable to obtain the spectral-weight
transfers with good qualitative agreement with the experimental results, thus
suggesting that such other contributions may be minor or irrelevant.

The inset of Fig. \ref{DeltaW_over_W}(a) is to show that for 0-1 eV region
our result does not agree with the experiment,
since it does not capture the insulator-metal transition.
As mentioned earlier, we argue that this is due to our model not
incorporating the dynamic Jahn-Teller phonons and their interactions with
electrons, which may be responsible to form an insulating gap in the
paramagnetic phase. The incorporation of such terms to improve our present
model is under our on-going study.

{\em{Conclusion.}} In conclusion, we have developed a model to explain
the structures and the spectral weight transfer occuring in the optical
conductivity of La$_{1-x}$Ca$_x$MnO$_3$ for $x=0.3$.
The key that makes our model work in
capturing the structures of the optical conductivity at medium and
high energies is the inclusion of O$_{2p}$ orbitals into the model.

Further, by parametrizing the hopping integrals through magnetization,
our model captures the spectral weight transfer as temperature
is decreased across the ferromagnetic transition temperature.
Our calculation based on this phenomenological parameters suggests that
the ferromagnatic ordering
increases the hopping parameter connecting the O$_{2p}$ orbitals and
the upper Mn$_{e_g}$ orbitals, while simultaneously
decreasing the hopping parameter
connecting O$_{2p}$ orbitals and the lower Mn$_{e_g}$ orbitals.
Although we have yet to check whether or not this scenario works
in a more complete model incorporating the dynamic electron-phonon coupling,
we conjecture that this may be of important part that contributes to
the mechanism of insulator to metal transition in manganites.

Overall, our results demonstrate the strength of our model that one may have
to consider as the minimum model before adding other ingredients
in order to properly explain the insulator-metal transition
or other features in correlated electron systems such as manganites.

{\em{Acknowledgement.}}
MAM and AR thank George Sawatzky and Seiji Yunoki for their valuable comments
and suggestions.
This work is supported by NRF-CRP grant "Tailoring Oxide Electronics by Atomic
Control" NRF2008NRF-CRP002-024, NUS YIA, NUS cross faculty grant and FRC.
We acknowledge the CSE-NUS computing centre for providing facilities for our
numerical calculations. Work at NTU was supported in part by a MOE AcRF
Tier-1 grant (grant no. M52070060).

\end{document}